\documentclass[manuscript]{aastex}

\usepackage{graphicx}
\usepackage{txfonts}
\usepackage{natbib}
\usepackage[caption=false]{subfig}
\usepackage{float}
\usepackage{bm}

\newcommand{\herschel}{\emph{Herschel}}
\newcommand{\cmc}{cm$^{-3}$}

\newcommand{\bsize}{$\theta_{b}$}
\newcommand{\trot}{$T_{\rm rot}$}
\newcommand{\tkin}{$T_{\rm kin}$}
\newcommand{\ntot}{$N_{\rm tot}$}

\newcommand{\eup}{$E_{\rm up}$}
\newcommand{\hh}{H$_{2}$}

\newcommand{\sn}{~$\times$~10}

\newcommand{\chisq}{$\chi^{2}$}

\newcommand{\kms}{km s$^{-1}$}
\newcommand{\tmax}{$T_{\rm max}$}
\newcommand{\tfinal}{$T_{\rm F}$}
\newcommand{\twarm}{$t_{\rm warm}$}

\newcommand{\nmols}{39}
\newcommand{\niso}{79}

\newcommand{\ncomplex}{8}

\shorttitle{analysis of the HIFI spectral survey toward Orion KL II}
\shortauthors{Crockett et al.}

\begin{document}
\title{\herschel\ observations of EXtra-Ordinary Sources: ANALYSIS OF THE HIFI 1.2~THz WIDE SPECTRAL SURVEY TOWARD ORION KL II. CHEMICAL IMPLICATIONS\footnote{Herschel is an ESA space observatory with science instruments provided by European-led Principal Investigator consortia and with important participation from NASA.}}

\author{N.~R. Crockett\altaffilmark{1}, E.~A. Bergin\altaffilmark{2}, J.~L. Neill\altaffilmark{2}, C. Favre\altaffilmark{2}, 
G.~A. Blake\altaffilmark{1}, E. Herbst\altaffilmark{3}, D.~E. Anderson\altaffilmark{1}, and G.~E. Hassel\altaffilmark{4}}

\altaffiltext{1}{California Institute of Technology, Division of Geological \& Planetary Sciences, MC 150-21, Pasadena, CA 91125, USA}
\altaffiltext{2}{Department of Astronomy, University of Michigan, 500 Church Street, Ann Arbor, MI 48109, USA}
\altaffiltext{3}{Departments of Chemistry and Astronomy, University of Virginia, Charlottesville, VA 22904, USA}
\altaffiltext{4}{Department of Physics and Astronomy, Siena College, Loudonville, NY 12211, USA}

\begin{abstract}
We present chemical implications arising from spectral models fit to the \herschel/HIFI spectral survey toward the Orion Kleinmann-Low nebula (Orion~KL). We focus our discussion on the \ncomplex\ complex organics detected within the HIFI survey utilizing a novel technique to identify those molecules emitting in the hottest gas. In particular, we find the complex nitrogen bearing species CH$_{3}$CN, C$_{2}$H$_{3}$CN, C$_{2}$H$_{5}$CN, and NH$_{2}$CHO systematically trace hotter gas than the oxygen bearing organics CH$_{3}$OH, C$_{2}$H$_{5}$OH, CH$_{3}$OCH$_{3}$, and CH$_{3}$OCHO, which do not contain nitrogen.  If these complex species form predominantly on grain surfaces, this may indicate N-bearing organics are more difficult to remove from grain surfaces than O-bearing species. Another possibility is that hot (\tkin~$\sim$~300~K) gas phase chemistry naturally produces higher complex cyanide abundances while suppressing the formation of O-bearing complex organics. We compare our derived rotation temperatures and molecular abundances to chemical models, which include gas-phase and grain surface pathways. Abundances for a majority of the detected complex organics can be reproduced over timescales $\gtrsim$~10$^{5}$~years, with several species being under predicted by less than 3$\sigma$. Derived rotation temperatures for most organics, furthermore, agree reasonably well with the predicted temperatures at peak abundance. We also find that sulfur bearing molecules which also contain oxygen (i.e. SO, SO$_{2}$, and OCS) tend to probe the hottest gas toward Orion~KL indicating the formation pathways for these species are most efficient at high temperatures.  
\end{abstract}
\keywords{astrochemistry --- ISM: abundances --- ISM: individual objects (Orion KL) --- ISM: molecules}

\section{Introduction}

Hot cores, the birth sites of massive stars, are the most prodigious emitters of molecular line radiation in the Galaxy, harboring high abundances of organic molecules \citep{herbst09}. Many of the organics detected in these regions are composed of 6 atoms or more, which are often labelled ``complex" by astronomers. Though the chemistry in these regions remains poorly understood, theoretical studies argue that the evaporation of icy grain mantles, precipitated by the formation of protostars, plays a key role in the production of many complex organic molecules \citep[see e.g.][]{millar91, charnley92, caselli93, charnley95, charnley97b, rodgers01, garrod06, garrod08}. However, the extent to which these species form on grain surfaces as opposed to in the gas phase remains unclear. 

The Orion Kleinmann-Low nebula (Orion~KL) is a high mass star forming region that provides a superb template for constraining the chemical processes active within these environments. Orion~KL harbors one of the most organic rich hot cores in the Milky Way \citep{blake87}. Moreover, the so-called compact ridge, a group of externally heated dense clumps residing $\sim$~12\arcsec\ south-west of the hot core \citep{wang11, favre11}, is equally rich in organic species. Relative to one another, the hot core and compact ridge are more abundant in organics that contain nitrogen and oxygen, respectively \citep{blake87, beuther05, guelin08}. This is a curious chemical difference, given the proximity of these two regions, which has motivated several theoretical studies \citep{millar91, caselli93, rodgers01}. High spatial resolution maps, which reveal substantial overlap between O and N-bearing organics (though the emission peaks are distinct), complicate this picture by showing the boundaries between O and N-bearing organic rich gas are not well defined \citep{friedel08, friedel12, weaver12}. This suggests that differences in the physical conditions (i.e. temperature, density, and gas kinematics) on small spatial scales play an important role in the complex morphology of organics in this region \citep[see e.g.][]{friedel12}. Additionally, Orion~KL harbors at least one outflow, often referred to as the plateau, which contains the complex organic CH$_{3}$CN \citep{wang10, bell14}. A detailed picture of the physical conditions from which each molecule emits toward Orion~KL and other high-mass star forming regions, therefore, provides critical constraints for theoretical studies that model the chemistry of complex organics in these environments. 

Unbiased spectral line surveys are useful tools for obtaining the chemical inventory of interstellar molecular gas and constraining the excitation of detected species. \citet[][hereafter Paper I]{crockett14a} presents a global analysis of the \herschel/HIFI 1.2~THz wide spectral survey of Orion~KL. These observations were taken as part of the guaranteed time program entitled \herschel\ \emph{Observations of EXtra Ordinary Sources} (HEXOS). Additional details concerning this program can be found in \citet{bergin10}. As part of the analysis, Paper~I presents models, which assume local thermodynamic equilibrium (LTE), of the molecular emission detected in the HIFI survey. These models provide estimates for the rotation temperature, \trot, and total column density, \ntot, (from which abundances are computed) of each species. In this follow up study, we discuss the chemical implications of the LTE models focusing on complex organics. 

This paper is organized as follows. In Section~\ref{s-obs}, we review the Orion~KL HIFI scan and LTE models. In Section~\ref{s-res}, we identify the molecules emitting in the hottest gas using a novel method which employs the emission predicted by our spectral models as a direct metric (\ref{s-emis}) and present gas-grain chemical models to help interpret our results (\ref{s-mod}). We discuss our results in Section~\ref{s-disc}, and present our conclusions in Section~\ref{s-con}.

\section{Observations and The Full Band Model}
\label{s-obs}

We obtained spectral scans in all available HIFI \citep{degraauw10} bands (1a -- 7b) between 1~March 2010 and 19~March 2011 using the wide band spectrometer (WBS). The full spectral survey thus covers a frequency range from 480 to 1907~GHz with two gaps at 1280 -- 1430~GHz and 1540 -- 1570~GHz. The WBS provides a uniform spectral resolution of 1.1~MHz (0.2 -- 0.7~\kms); the different spatial/velocity components associated with Orion KL are thus easily resolved by HIFI. All of the observations were taken in dual beam switch (DBS) mode. At HIFI frequencies, the beam size, \bsize, of \herschel\ varies between 11\arcsec\ and 44\arcsec. In bands 1 -- 5 (\bsize~=~17\arcsec -- 44\arcsec) the telescope was pointed toward coordinates $\alpha_{J2000} = 5^h35^m14.3^s$, $\delta_{J2000} = -5^{\circ}22'33.7''$, midway between the hot core and compact ridge. In bands 6 and 7, because of the smaller beam size (\bsize~=~11\arcsec\ -- 15\arcsec), separate observations were obtained toward the hot core ($\alpha_{J2000} = 5^h35^m14.5^s$, $\delta_{J2000} = -5^{\circ}22'30.9''$)  and compact ridge ($\alpha_{J2000} = 5^h35^m14.1^s$, $\delta_{J2000} = -5^{\circ}22'36.5''$). The data were reduced using the standard HIPE \citep{ott10} pipeline (version 5.0, build 1648), which produces ``Level 2" flux calibrated double side band (DSB) spectra. HIFI provides separate horizontal (H) and vertical (V) polarizations.  The DSB spectra for both polarizations were deconvolved using the \emph{doDeconvolution} task within HIPE to create single sideband (SSB) spectra \citep{comito02} after the DSB scans were baseline subtracted and spurious spectral features, not flagged by the pipeline, were removed. As a final step, we averaged the H and V polarizations together. 

We modeled the molecular emission using the XCLASS\footnote{http://www.astro.uni-koeln.de/projects/schilke/XCLASS} program. In total, we detect emission and/or absorption from \niso\ isotopologues originating from \nmols\ molecules, most of which have their own molecular fits (those species observed predominantly in absorption or with only a few detected transitions over a limited range in excitation energy were not fit). Summing the predicted emission from all detected species yields the ``full band model". The fits are constrained at mm wavelengths by combining the HIFI spectrum with a ground-based line survey obtained with the IRAM 30~m telescope \citep{tercero10}. This survey spans frequency ranges of 80 -- 116~GHz, 130 -- 180~GHz, and 200 -- 280~GHz, corresponding to atmospheric windows at 3~mm, 2~mm, and 1.3~mm, respectively. The models are thus constrained from the mm to the far-IR and, in almost all cases, include transitions at or close to ground up to excitation energies where emission is no longer detected. A reduced \chisq\ analysis shows generally excellent agreement between the data and models. This is particularly true for complex organics, which typically have reduced \chisq\ values $\le$~1.5 and were constrained by hundreds of lines detected at or above the 3~$\sigma$ level. 

The full Orion~KL HIFI spectral survey along with a more detailed description of the data reduction procedure and modeling methodology are presented in Paper~I and  \citet{crockett14b}. In this study, we use the molecular fits as templates for the data, analyzing the emission from the models rather than the data. Examining the HIFI scan in this way has two advantages. First, we are able to easily separate emission from different spatial/velocity components. And second, we do not have to be concerned with line blends because we examine the model emission on a per molecule basis.

\section{Results}
\label{s-res}

\subsection{Characterizing the Molecular Emission}
\label{s-emis}

\begin{figure}
\epsscale{0.9}
\plotone{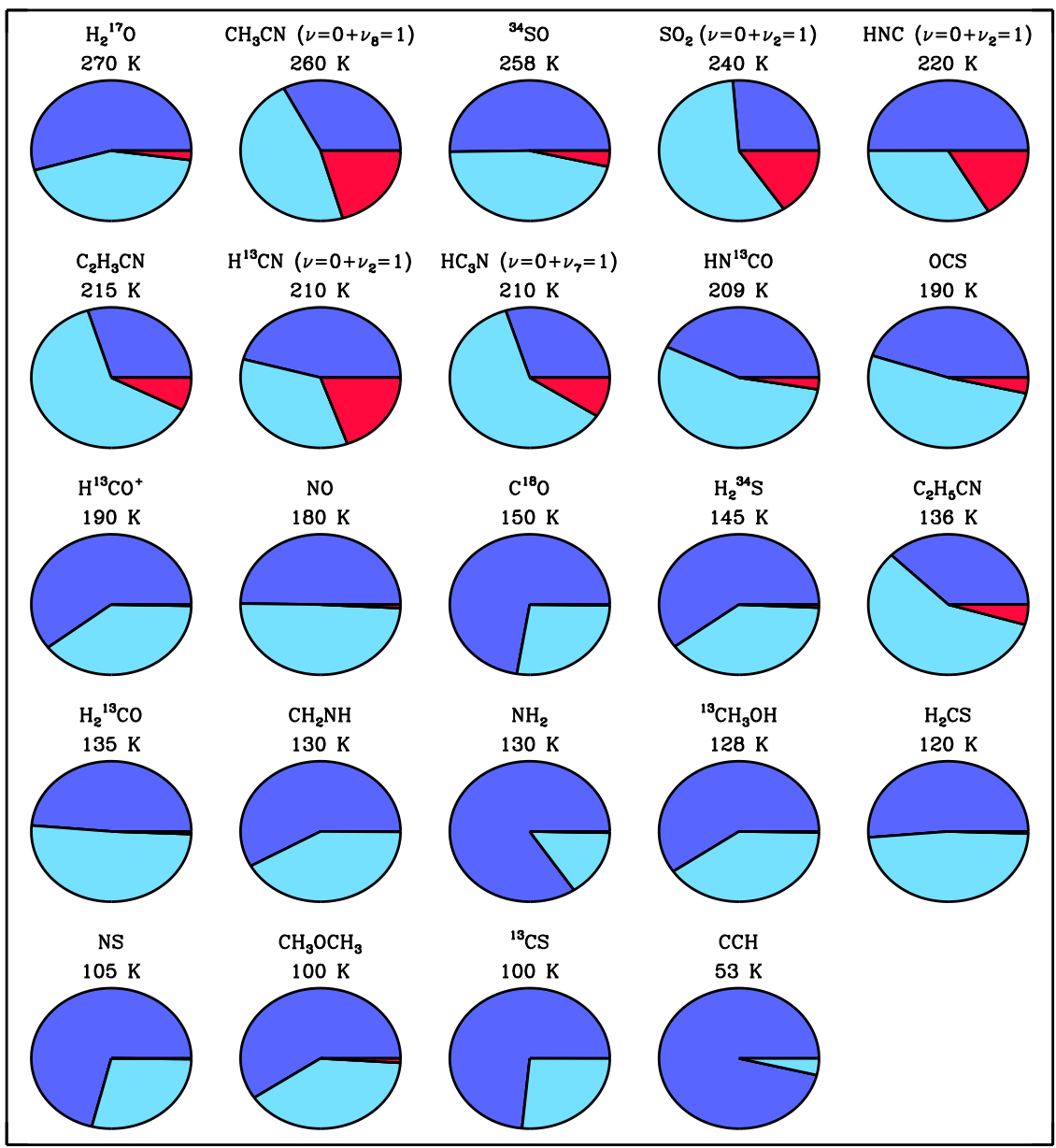}
\caption{Pie charts that plot the fraction of total integrated intensity originating from states in \eup\ ranges 0 -- 200~K (dark blue), 200 -- 800~K (cyan), and 800 -- 3000~K (red) for molecules detected toward the hot core. The molecule ID and \trot\ derived in Paper I are given above each chart. If a chart includes emission from a vibrationally excited state, it is indicated in parentheses with the molecule ID. \label{p-hc}}
\end{figure}

\begin{figure}
\epsscale{0.9}
\plotone{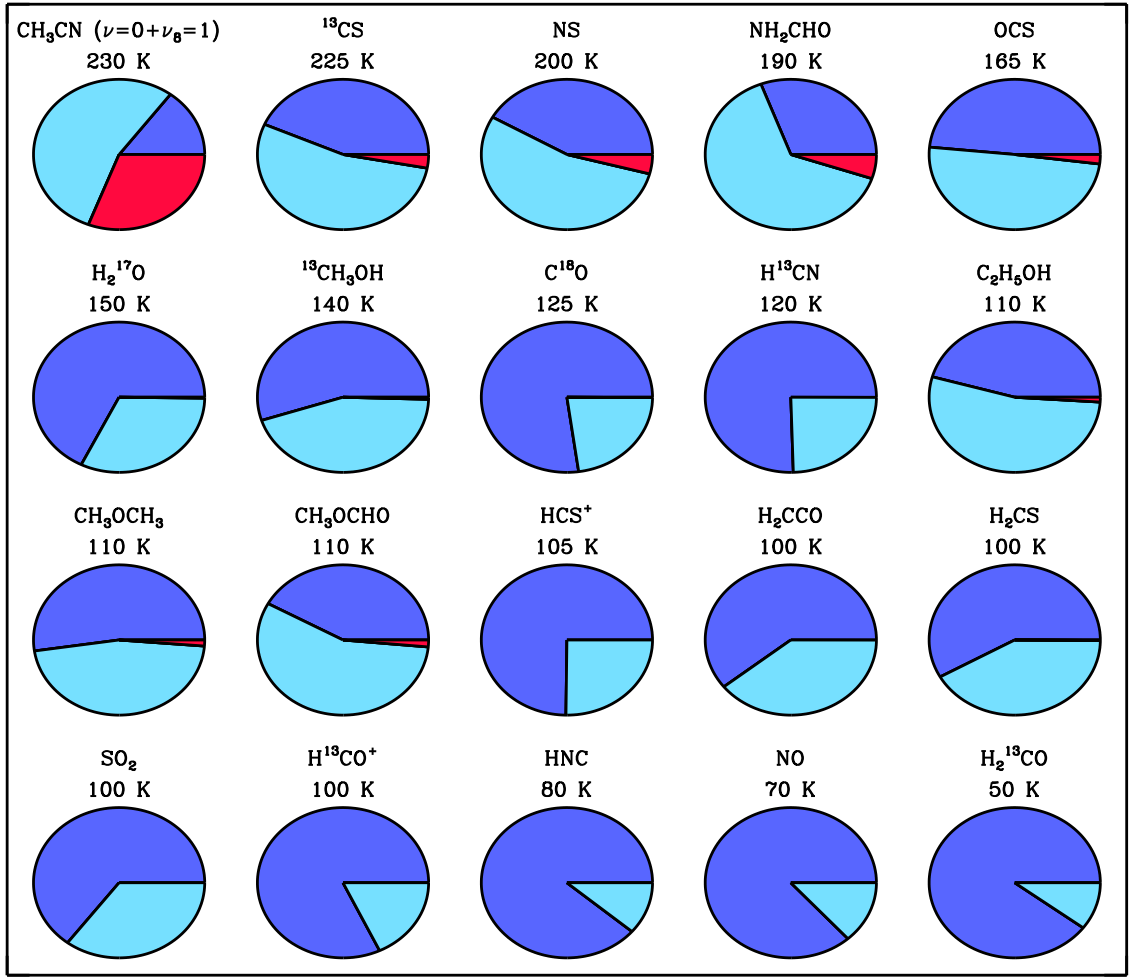}
\caption{Same as Fig.~\ref{p-hc} for the compact ridge. \label{p-cr}}
\end{figure}

\begin{figure}
\epsscale{0.9}
\plotone{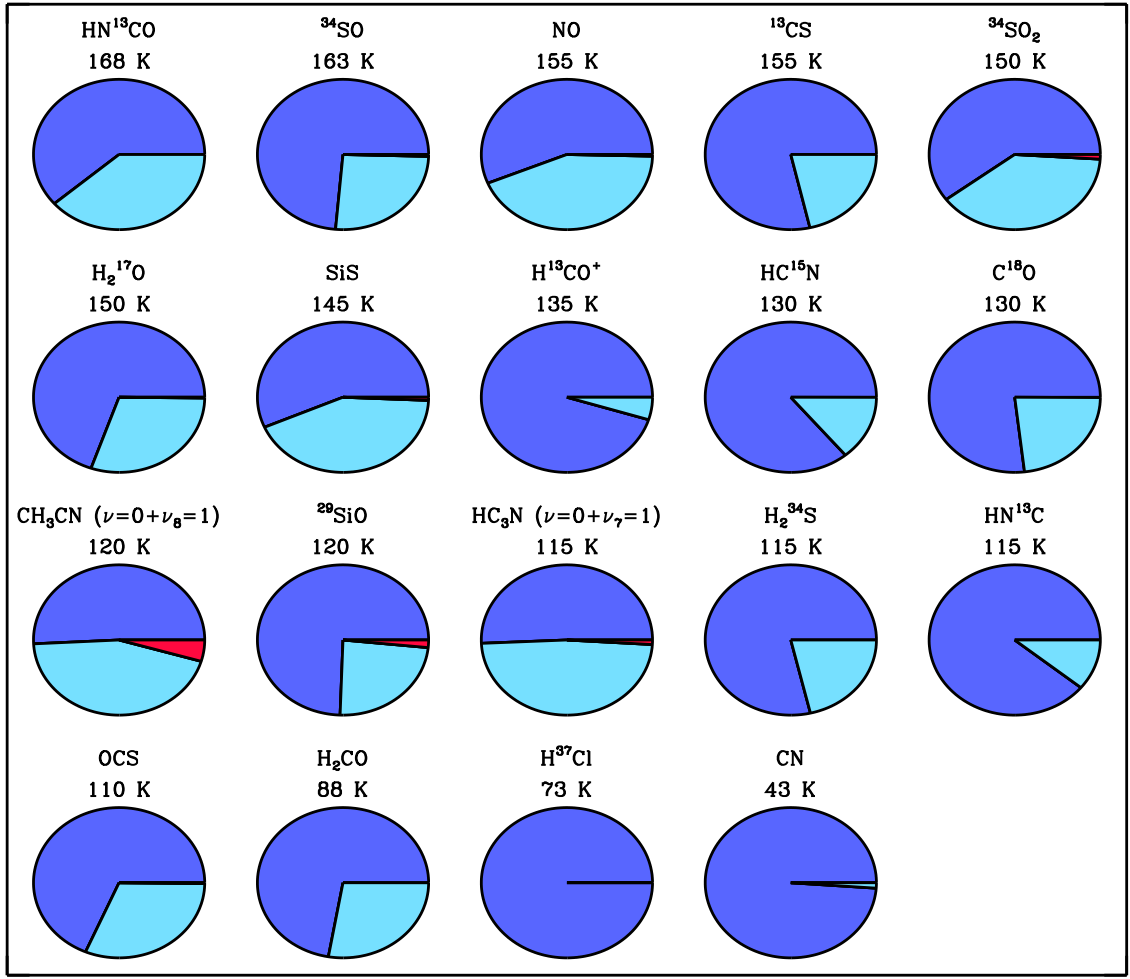}
\caption{Same as Fig.~\ref{p-hc} for the plateau. \label{p-pl}}
\end{figure}

We employ a novel diagnostic to characterize the molecular line emission within the Orion~KL HIFI scan. For each molecule and spatial/velocity component, we sum the integrated intensity from all \trot\ sub-components presented in Paper~I. We also include emission from  vibrationally excited modes if they are detected for a given species. We then compute the fraction of total integrated intensity originating from states over different ranges of upper state energy, \eup. Figs.~\ref{p-hc}, \ref{p-cr}, and \ref{p-pl} plot these fractions as pie charts for molecules detected toward the hot core, compact ridge, and plateau, respectively. We do not discuss emission from the extended ridge here because no complex organics are detected toward this component. Molecule names and rotation temperatures derived in Paper~I are given above each pie chart. If a pie chart includes a vibrationally excited mode, it is also indicated in the molecule label. The charts are organized so that \trot\ decreases from the upper left to the lower right.  In instances when more than one isotopologue is detected for a given molecule, we represent the emission using a rarer species to avoid issues with high optical depth. However, when vibrationally excited modes are only detected in more abundant isotopologues (e.g. CH$_{3}$CN, SO$_{2}$, etc.), we adopt the more abundant species so that emission from vibrationally excited states can be included in the pie charts. For the sake of simplicity, we use only main isotopologue names in the discussion below even if the pie chart for a particular molecule is made using the spectrum of a rarer species.

The above metric is a more comprehensive diagnostic for the excitation of a given molecule compared to a single \trot\ because many species had to be fit with multiple temperature sub-components indicating the presence of temperature gradients. Such molecules therefore probe a range of temperature environments within a given spatial/velocity component. Moreover, in order to accurately fit the vibrationally excited emission of some species, we had to adjust \trot, and, in some instances, increase \ntot\ relative to the ground vibrational state, suggesting these molecules are being pumped from mid/far-IR photons.

Examining Figs.~\ref{p-hc} -- \ref{p-pl}, we see that those molecules emitting a significant fraction of their total line radiation from high energy states (\eup~$>$~800~K) are readily apparent. Such molecules represent the most highly excited molecules within Orion KL and likely originate from the hottest gas. We emphasize that this metric is independent of rotation temperature and relies solely on the observed emission as predicted by our models. Ethyl cyanide (C$_{2}$H$_{5}$CN) is the only cyanide detected toward the hot core with a rotation temperature less than 200~K (\trot~=~136~K). Its LTE model, however, requires a hotter, more compact sub-component (\trot~=~300~K) to accurately fit the emission (Paper I). The presence of this sub-component is evidenced in the pie chart of C$_{2}$H$_{5}$CN because approximately 5\% of its integrated intensity originates from states with \eup~$>$~800~K. This is in contrast to molecules with similar rotation temperatures (e.g. H$_{2}$CO), which do not emit significantly from states with \eup~$>$~800~K. Methyl cyanide (CH$_{3}$CN), furthermore, does not have an especially high \trot\ toward the plateau suggesting it does not probe particularly hot gas in the outflow. Including emission from the $\nu_{8}$~=~1 vibrationally excited mode (\eup~$\ge$~520~K), however, indicates that it may be probing hotter material than suggested by the derived \trot.

Of the \nmols\ molecules detected in the HIFI spectrum, \ncomplex\ are complex. This subset includes species which contain only carbon, hydrogen, and oxygen as well as nitrogen bearing organics which may also contain oxygen (i.e. NH$_{2}$CHO). From Fig.~\ref{p-hc}, we see that among the 5 complex molecules detected in the hot core, the nitrogen bearing species CH$_{3}$CN, C$_{2}$H$_{3}$CN, and C$_{2}$H$_{5}$CN, all of which are cyanides, display more highly excited emission, as represented by the pie charts, than their oxygen bearing counterparts, CH$_{3}$OH and CH$_{3}$OCH$_{3}$. Moreover, the simpler cyanides HCN, HNC, and HC$_{3}$N also display some of the most highly excited emission toward the hot core. Fig.~\ref{p-cr} shows a similar trend in the compact ridge. The N-bearing complex organics CH$_{3}$CN and NH$_{2}$CHO display more highly excited emission than organics which contain only C, H, and O: CH$_{3}$OH, C$_{2}$H$_{5}$OH, CH$_{3}$OCH$_{3}$, and CH$_{3}$OCHO. Finally, Fig.~\ref{p-pl} shows that CH$_{3}$CN, the only complex organic detected toward the plateau, displays the most highly excited spectrum. 

Figs.~\ref{p-hc} -- \ref{p-pl} also show that sulfur bearing molecules which also contain oxygen, i.e. SO, SO$_{2}$, and OCS, make up another subset of molecules that trace the hottest gas toward the hot core, compact ridge, and plateau. The two exceptions to this are SO$_{2}$ detected toward the compact ridge (\trot~=~100~K) and OCS detected toward the plateau (\trot~=~110~K). We note, the rotation temperature of SO$_{2}$ toward the compact ridge is not well constrained because the line profiles are dominated by the plateau and hot core and thus could be underestimated. The relatively low \trot\ value for OCS, combined with its pie chart, however, suggest it is probing cooler material than SO and SO$_{2}$ toward the plateau. 

\subsection{Chemical Modeling}
\label{s-mod}

In order to determine if theoretical studies agree with our derived abundances and \trot\ values, we compare our results to a grid of four chemical models using the OSU gas-grain chemical code \citep{garrod06, garrod07, garrod08}. 
We note that more complex chemical models exist which include 1-D hydrodynamical treatments \citep[e.g.][]{aikawa12}. These models, however, are not easily compared with our single dish results because the predicted molecular abundances depend upon a time dependent physical profile (i.e. temperature, density, and velocity field as a function of radius). Because we do not attempt to model the physical structure or dynamical history of Orion KL here, we use the simpler ``warm-up" approach described below.
We emphasize that a global chemical model that reproduces the abundances of all observed species in the Orion~KL HIFI survey is beyond the scope of this work because predicted gas phase abundances are sensitive to the details of the thermal history for a given region, binding and desorption energies of grain surface molecules, and reaction branching ratios, all of which can be highly uncertain. Rather, we aim here to determine if the OSU gas-grain chemical code, which assumes a relatively simple thermal history (see below), predicts abundances for complex organics that are comparable to what is observed in the HIFI scan and if this model can reproduce the observed trend of N-bearing complex molecules probing hotter gas relative to O-bearing organics. 

The OSU gas-grain chemical model assumes a two phase evolution. The first phase is an isothermal collapse in which the gas density increases from 3\sn$^{3}$ to 1\sn$^{7}$~\cmc\ at a temperature of 10~K over $\sim$10$^{6}$~years. The second is a ``warm-up" phase in which the gas temperature increases by following a quadratic power law over a warm-up timescale, \twarm, to a final temperature, \tfinal. The gas and dust are assumed to be well coupled to one another, thus the temperatures of both components are assumed to be equal. Within this model, complex organics are formed both in the gas phase and on grain surfaces. Many of the complex organics are produced by reactions between radicals, which are formed primarily by photodissociating molecules abundant in ice mantles, during the warm-up phase. Because the warm-up stage is characterized by high dust extinction, most of the photodissociation is the result of photons induced by cosmic rays \citep{ruffle01}. The smooth rise in temperature over time is therefore key because it allows heavy radicals to become mobile on grain surfaces without being sublimated, and facilitates gas phase pathways with activation energies. For the grid, we adopt two values for \tfinal\ of 200~K and 300~K, commensurate with the highest \trot\ values measured in the HIFI survey, and two values for \twarm\ of 5\sn$^{4}$~years and 2\sn$^{5}$~years, corresponding to high and intermediate mass star formation as laid out by \cite{viti04}. All other model parameters, including the chemical network, are the same as \citet{garrod08}. We also adopt the same initial fractional abundances from the ``standard" model of \citet{garrod08}.

\begin{figure}
\epsscale{1.09}
\plotone{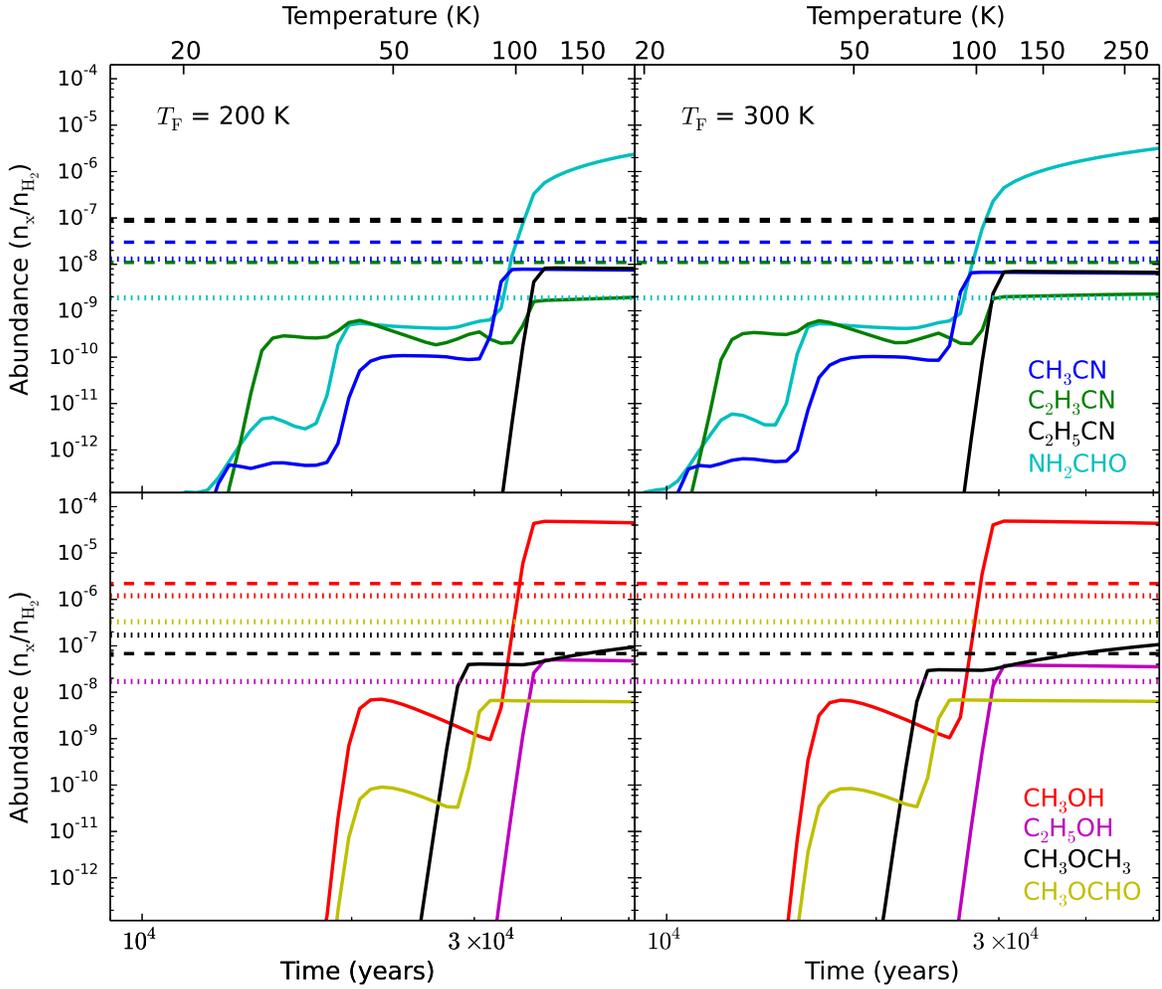}
\caption{Predicted gas phase abundances with respect to \hh\ for the eight complex organics detected in the Orion~KL HIFI survey plotted as a function of time for \twarm~=~5~\sn$^{4}$~years. The abundance of each species is plotted as a different color (solid lines). Abundances for N and O-bearing organics are plotted in the top and bottom rows, and models with \tfinal~=~200 and 300~K are plotted in the left and right columns, both respectively. Observed abundances toward the hot core and compact ridge are plotted as horizontal dashed and dotted lines, respectively, with the same color scheme as the model values. The bottom x-axis is time such that zero corresponds to the beginning of the warm-up phase and the top x-axis displays the corresponding temperature.
 \label{p-modf}}
\end{figure}

\begin{figure}
\epsscale{1.09}
\plotone{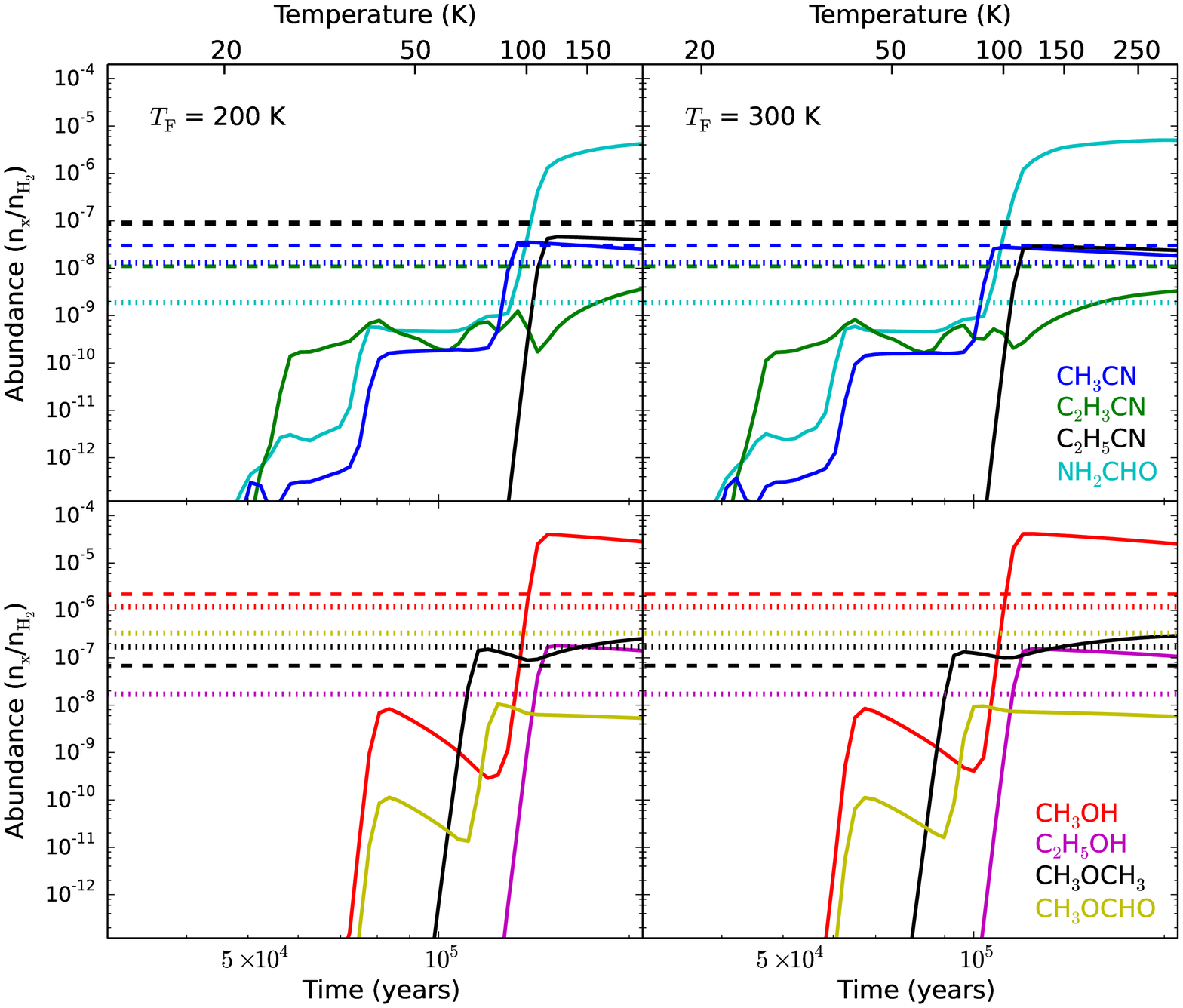}
\caption{Same as Figure~\ref{p-modf} for \twarm~=~2\sn$^{5}$~years. \label{p-modm}}
\end{figure}

Figures~\ref{p-modf} and \ref{p-modm} plot the predicted gas phase abundances of the eight complex molecules detected in the HIFI survey as a function of time during the warm-up phase for \twarm~=~5~\sn$^{4}$~years and 2~\sn$^{5}$~years, respectively. Measured abundances for these species toward the hot core and compact ridge (Paper~I) are also overlaid as horizontal lines. Table~\ref{t-obs} reports the temperature at which the abundance of each molecule is maximized, \tmax, along with the peak abundance. Examining Table~\ref{t-obs}, we see that changing \tfinal\ or \twarm\ does not strongly affect \tmax, which varies by $\lesssim$~11\% except when a molecule's peak abundance occurs at or just before \tfinal, in which case \tmax~$\approx$~\tfinal. The gas-phase abundances of those species with \tmax~$<$~\tfinal\ typically peak at temperatures somewhat higher than their sublimation temperature when they thermally desorb from grains, which is why \tmax\ does not vary much for these molecules. The efficient sublimation of a molecule is shown in Figures~\ref{p-modf} and \ref{p-modm} as a steep rise in gas phase abundance followed by a slow decrease corresponding to destruction pathways which are active at high temperatures. We refer here to rises in abundance that occur last in the evolution of the models, but note additional earlier abundance peaks occur for many of the organics which are the result of interdependent gas phase and grain surface pathways. Those species with \tmax~$\approx$~\tfinal, on the other hand, (i.e. C$_{2}$H$_{3}$CN, NH$_{2}$CHO, and CH$_{3}$OCH$_{3}$) have formation/destruction pathways which yield net increases in their abundances after sublimation from grain surfaces. The peak abundance varies more significantly than \tmax\ across the grid (by as much as a factor of $\sim$~7), especially when \twarm\ is changed. As noted previously, the OSU code typically predicts higher abundances for complex organics when \twarm\ is longer because there is more time at intermediate temperatures ($\sim$~20 -- 40~K) when heavy radicals are mobile on grain surfaces \citep{garrod06, garrod08}. The parameter \twarm\ thus plays the more important role relative to \tfinal\ in the production of complex molecules.

We include in our analysis the N-bearing organics C$_{2}$H$_{3}$CN and C$_{2}$H$_{5}$CN which have not been discussed in earlier studies that employ the OSU chemical network. Within the model, C$_{2}$H$_{3}$CN forms predominantly in the gas phase via the reaction,
\begin{equation}
\rm CN + C_{2}H_{4} \to C_{2}H_{3}CN + H,
\end{equation}
though dissociative recombination of C$_{3}$H$_{4}$N$^{+}$ contributes somewhat when temperatures are $\lesssim$~40~K. The main destruction routes are through the ion-neutral reaction,
\begin{equation}
\rm HCO^{+} + C_{2}H_{3}CN \to C_{3}H_{4}N^{+} + CO,
\end{equation}
and accretion onto dust grains. At temperatures $\gtrsim$~40~K accretion onto dust grains is the major destruction mechanism for gas-phase C$_{2}$H$_{3}$CN, after which C$_{2}$H$_{3}$CN is hydrogenated on grain surfaces (\tkin~$\lesssim$~120~K) or quickly thermally desorbed back into the gas phase (\tkin~$\gtrsim$~120~K). The formation of C$_{2}$H$_{5}$CN occurs on grain surfaces via multiple hydrogenations of HC$_{3}$N and C$_{2}$H$_{3}$CN. C$_{2}$H$_{5}$CN remains on grain surfaces until it is efficiently thermally sublimated at temperatures $\gtrsim$~100~K, at which point the gas phase abundance rises rapidly.

Comparing the model predictions with observations, we see that the chemical models reproduce observed abundance levels for a majority of the complex organics at some point in time within the grid.  The species C$_{2}$H$_{3}$CN, C$_{2}$H$_{5}$CN, and CH$_{3}$OCHO, however, are consistently under predicted, though the agreement is better for the longer \twarm, which produce systematically higher abundances for these species than the shorter \twarm. We note that the abundances for these molecules are under predicted by less than 3$\sigma$ for \twarm~=~2\sn$^{5}$~years given the uncertainties in the observations (Paper~I). Moreover, abundances for CH$_{3}$CN, toward the hot core and compact ridge, as well as CH$_{3}$OCH$_{3}$, toward the compact ridge, are only under predicted for \twarm~=~5\sn$^{4}$~years. Hence, we conclude models corresponding to \twarm~=~ 2~\sn$^{5}$~years, in general, produce abundances that agree more closely with the observed values. For these models, observed abundances are reached at times $\gtrsim$~10$^{5}$~years. We note that our model grid predicts peak abundances for NH$_{2}$CHO that are more than three orders of magnitude higher than that observed toward the compact ridge and even approximately 2 orders of magnitude higher than the NH$_{2}$CHO abundance measured toward SagB2(N) \citep{neill14}. This overabundance is a result of the fact that NH$_{2}$CHO forms efficiently in the gas phase by the combination of NH$_{2}$ and H$_{2}$CO, and is continually accreted onto and desorbed from dust grains without an efficient destruction mechanism late in the model's evolution. Currently the major gas-phase destruction mechanism (behind accretion onto dust grains) is via reactions with H$_{3}$O$^{+}$ and H$_{3}^{+}$ that form NH$_{2}$CH$_{2}$O$^{+}$, which is subsequently destroyed by dissociative recombination producing NH$_{2}$, HCO, and H$_{2}$CO. Although the gas phase route helps to explain the presence of NH$_{2}$CHO in hot gas, we emphasize that its efficiency is quite uncertain and likely too large to explain abundances observed toward hot cores.

\section{Discussion}
\label{s-disc}

\subsection{Chemical Implications}
Our results indicate that complex N-bearing organics, i.e. cyanides and NH$_{2}$CHO, probe hotter environments than complex species which contain no nitrogen. Previous single dish and interferometric studies have shown that gas which traces the hot core is hotter than gas which traces the compact ridge \citep[see e.g.][]{blake87, beuther05}. Because the hot core is richer in N-bearing organics, these molecules tend to trace hotter gas than their O-bearing counterparts. Recently \citet{friedel12} overlaid high spatial resolution maps of key N/O-bearing complex molecules with a temperature map constructed from a combination of their own CH$_{3}$OH data and NH$_{3}$ observations from \citet{goddi11}. They showed that the N-bearing species C$_{2}$H$_{5}$CN, C$_{2}$H$_{3}$CN, and CH$_{3}$CN are cospatial with one another and trace higher temperature regions associated with the hot core and nearby sources SMA1 and IRc7. The O-bearing molecules CH$_{3}$OCHO and CH$_{3}$OCH$_{3}$, also cospatial with one another \citep[see also][]{favre11b, brouillet13}, have emission peaks near cooler regions associated with the compact ridge and sources IRc5 and IRc6. We emphasize that our results apply to molecules detected within a spatial/velocity component, i.e. N-bearing organics within the hot core are hotter relative to O-bearing molecules within the hot core with the same being true for the compact ridge. Our excitation temperatures, furthermore, are derived individually from the molecular emission of each species, and are the most robust to date because the emission is constrained by hundreds to thousands of transitions over mm, sub-mm, and far-IR wavelengths. The inclusion of up to thousands of lines per molecule in our analysis thus places these results on a strong statistical footing.

If both O and N-bearing complex organics form predominantly on grain surfaces, cyanides along with NH$_{2}$CHO may be more difficult to remove from grain surfaces than O-bearing species. Within a hot core, we might then expect oxygen bearing organics to be released during an earlier, presumably cooler, epoch and/or further from the central protostar. Measured excitation temperatures should thus be higher toward the same location, and spatial temperature variations larger and clumpier for N-bearing organics relative to O-bearing species because the former traces hotter material along the line of sight. Because interferometric observations show Orion~KL has structure on spatial scales $<$~1\arcsec\ \citep[e.g.][]{wang10, goddi11, favre11, friedel11}, our results illustrate the need for excitation temperature maps derived from both O and N-bearing organics at sub-arcsecond resolution to see if differences exist. Such observations will surely be attainable with ALMA. Existing interferometric maps of the kinetic temperature derived from CH$_{3}$CN \citep{wang10} and CH$_{3}$OH \citep{friedel12} tentatively support this viewpoint at least toward the compact ridge where \trot\ values are measured to be $\sim$~170 -- 280~K for CH$_{3}$CN and $\lesssim$~200~K for CH$_{3}$OH. We, however, note that the spatial resolution of the CH$_{3}$CN observations (1.6\arcsec~$\times$~0.9\arcsec) is about a factor of two higher relative to the CH$_{3}$OH map (2.2\arcsec~$\times$~2.0\arcsec). 

Another possibility is that hot gas phase chemistry may be producing the highly excited cyanides in the hot core. \citet{wang10}, for example, argue that CH$_{3}$CN in the hot core likely forms in the gas based on the chemical models of \citet{rodgers01}. These models show that at high temperatures (\tkin~=~300~K), hot cores produce CH$_{3}$CN, HC$_{3}$N and HCN in the gas phase at observed abundance levels on timescales of $\lesssim$~10$^{5}$ -- 10$^{6}$~years. Models presented by \citet{caselli93} show similar results. In this scenario, cyanides naturally trace hotter material because they form efficiently in the gas phase at high temperatures. \citet{hernandez14}, furthermore, presents a correlation in the abundance of CH$_{3}$CN with temperature based on LTE modeling of the 12$_{K}$ -- 11$_{K}$ ladder toward 17 hot cores, which suggests hotter material facilitates methyl cyanide formation, presumably in the gas phase, because measured temperatures in the hottest component of each model (2 temperature components were used for each core) range from approximately 100 -- 500~K. \citet{gerin92} reports a CH$_{2}$DCN~/~CH$_{3}$CN ratio $\ge$~0.005 toward the Orion~KL hot core, commensurate with D/H ratios measured for H$_{2}$O \citep{neill13a} and CH$_{3}$OH \citep{neill13b}, suggesting at least some methyl cyanide forms via grain surface chemistry at low temperatures. If gas phase formation routes are active for N-bearing organics in the hottest gas, we would expect gradients in the D/H ratios of complex N-bearing species like methyl cyanide. 

Previous observational studies have investigated possible differentiation between O and N-bearing organics toward other hot cores. Using single dish spectroscopic observations in the mm and sub-mm, \citet{bisschop07} computed the abundance and rotation temperatures for several O and N-bearing organics toward a sample of 7 hot cores. They found that the abundances between the complex ``hot", which they define as \trot~$>$~100~K, O-bearing species CH$_{3}$OCH$_{3}$, CH$_{3}$OH, C$_{2}$H$_{5}$OH, CH$_{3}$OCHO, and H$_{2}$CO are correlated with one another. However, abundances between the O-bearing and the N-bearing molecules HNCO and NH$_{2}$CHO, both of which trace hotter gas than the O-rich organics toward Orion KL (Fig.~\ref{p-hc} and \ref{p-cr}, respectively), did not correlate. \citet{bisschop07} argue that these results indicate that complex O and N-bearing organics originate from grain surface chemistries that are not strongly coupled to one another. Conversely, \citet{fontani07} carried out a similar study focusing on C$_{2}$H$_{5}$CN, C$_{2}$H$_{3}$CN, and CH$_{3}$OCH$_{3}$ using a sample of 6 hot cores. They found that abundances for all three species correlated with one another indicating that a simple oxygen/nitrogen chemical dichotomy is likely an over simplification. Indeed several interferometric studies have noted that (CH$_{3}$)$_{2}$CO has a unique morphology toward Orion~KL, peaking primarily toward regions where emission from both O and N-bearing complex organics are detected \citep{friedel08, friedel12, weaver12}. C$_{2}$H$_{5}$OH also has a unique morphology toward Orion~KL peaking at or near the (CH$_{3}$)$_{2}$CO peak as opposed to tracing CH$_{3}$OCHO and CH$_{3}$OCH$_{3}$ \citep{guelin08}. Many of these studies have also noted that there is significant overlap in the spatial distributions of large organics that contain O and N, indicating that chemically differentiation toward Orion~KL, in the strictest sense, is too elementary a picture. 

Our results also show that S~$+$~O-bearing molecules probe environments as hot as N-bearing complex organics, suggesting the formation pathways for these species favor high densities and temperatures. Most theoretical studies which trace sulfur chemistry in hot cores assume H$_{2}$S forms on grain surfaces via hydrogenation of S atoms \citep{millar97, charnley97, hatchell98}. Within these models, all other S-bearing molecules are produced in the gas phase after H$_{2}$S is evaporated from grain surfaces. Once in the gas, H$_{2}$S is efficiently destroyed by H atoms or other radicals or ions such as H$_{3}$O$^{+}$ with most of the sulfur eventually channeled into SO and SO$_{2}$ through reactions with OH and atomic O. Paper~I reports H$_{2}$S hot core abundances that are higher than that of SO and SO$_{2}$ by factors of 20 and 8, respectively, suggesting H$_{2}$S is being actively replenished via evaporation from icy grain mantles. The high rotation temperatures of SO, SO$_{2}$, and OCS, thus, may mean that  sulfur is being most efficiently channeled from H$_{2}$S into these species within the hottest gas. The fact that OCS traces cooler gas than SO and SO$_{2}$ toward the plateau, however, may indicate a unique formation route within the outflow for this species, possibly related to shock chemistry. 

\subsection{Comparison with Chemical Models}
Do the models presented in Section~\ref{s-mod}, which assume a rather simple warm-up phase, predict that N-bearing organics should systematically probe hotter gas than O-bearing organics? We take the same approach as \citet{garrod08} and compare the observed \trot\ to the predicted \tmax\ (Table~\ref{t-obs}). As noted in \citet{garrod08}, the OSU chemical code computes abundances for a single point in space over time. However, molecules with modeled abundances that peak at high temperatures represent better tracers of hot gas. Orion KL is a complex region with temperature and density gradients. Different species probe different environments, with the hottest and densest gas corresponding to material in a more evolved stage in its warm-up progression relative to cooler, more extended material. The comparison of \trot\ with \tmax\ is thus an efficient way to determine if the chemical models predict N-bearing organics are tracers of the hottest gas.

\begin{figure}
\epsscale{0.92}
\plotone{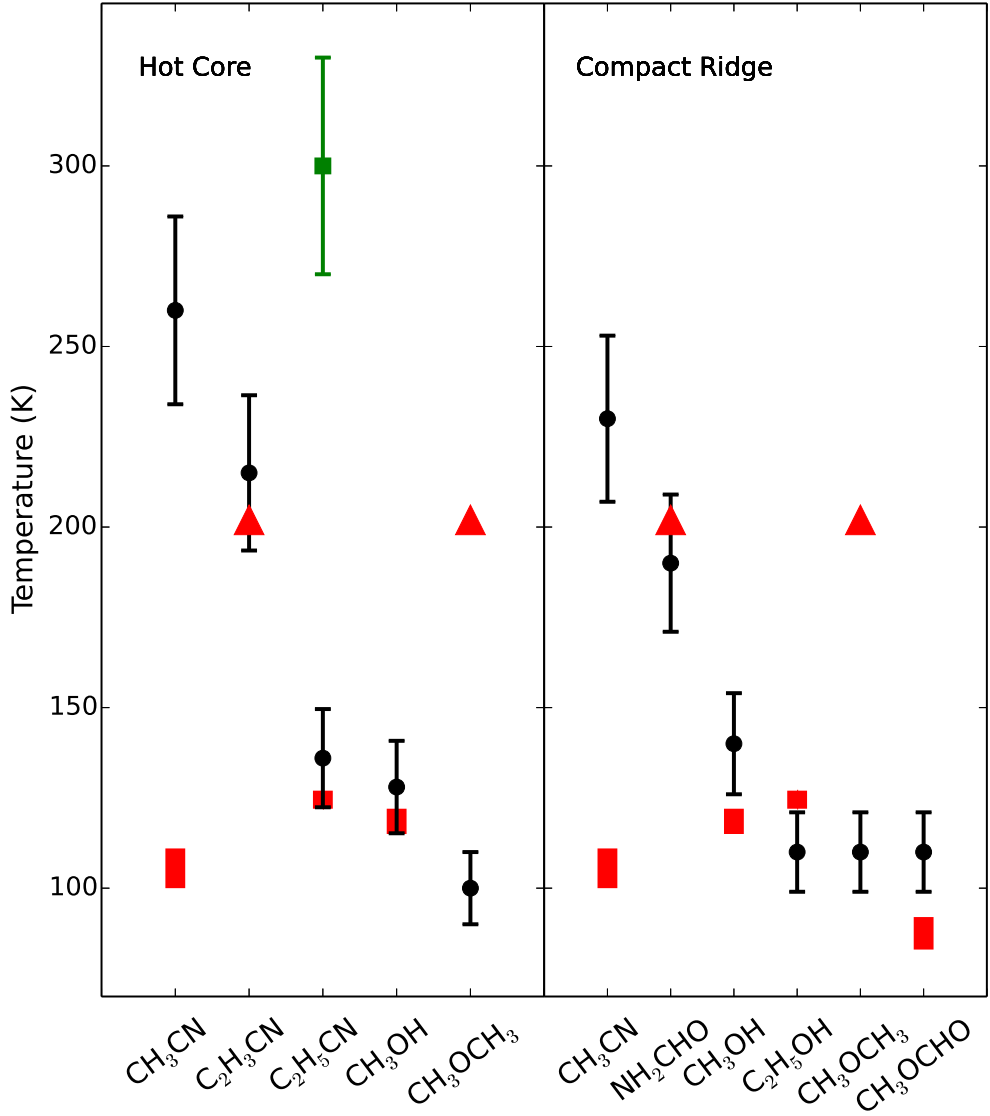}
\caption{Rotation temperatures (black dots) measured toward the hot core (left panel) and compact ridge (right panel) plotted as a function of molecule with \tmax\ values from our model grid (Table~\ref{t-obs}) overlaid. The range of \tmax\ predicted over all grid models for each molecule is represented by a red bar if \tmax~$<$~\tfinal\ and a red triangle at 200~K if \tmax~$\approx$~\tfinal\ indicating these molecules probe temperatures $\ge$~200~K in our model grid (see Section~\ref{s-mod}). The green square represents the hot, compact component required to fit the C$_{2}$H$_{5}$CN emission (see Section~\ref{s-emis}) in the HIFI scan. The error bars on the observed \trot\ values represent 1$\sigma$ uncertainties. \label{p-temp}}
\end{figure}

Figure~\ref{p-temp} plots \trot\ for the complex organics detected in the HIFI survey with \tmax\ values overlaid as a function of molecule toward the hot core and compact ridge. For C$_{2}$H$_{5}$CN, we plot the \trot\ value of the compact 300~K component required to fit the emission (see Sec.~\ref{s-emis}) as a green square to indicate that this species probes gas as hot as the other cyanides toward the hot core. We note that abundances plotted in Figures~\ref{p-modf} and \ref{p-modm} for most species stay relatively flat late in their evolution, which yield curves that do not have sharp peaks. Consequently, these molecules are predicted to be abundant over a range of temperatures, but are most abundant either right after sublimation (i.e. CH$_{3}$CN, C$_{2}$H$_{5}$CN, CH$_{3}$OH, C$_{2}$H$_{5}$OH, and CH$_{3}$OCHO) or continue to rise in abundance as the gas gets hotter (i.e. C$_{2}$H$_{3}$CN, NH$_{2}$CHO, and CH$_{3}$OCH$_{3}$). Our values for \tmax\ are thus not meant to be robust predictions for \trot. However, because the high molecular abundances observed toward Orion~KL are primarily due to active grain mantle sublimation, we use \tmax\ as a metric to, first, identify which molecules are predicted to be most abundant in the hottest gas, second, determine if there is general agreement between the observations and our models, and, third, identify the molecules that are obviously discrepant. 

Figure~\ref{p-temp} shows that most species have \tmax\ / \trot\ combinations that agree to within 2$\sigma$. However, CH$_{3}$CN and CH$_{3}$OCH$_{3}$ have \tmax\ predictions that are markedly different from their corresponding \trot\ values. For CH$_{3}$CN, we observe \trot~$\gtrsim$~230~K, but the models yield \tmax~$\leq$~111~K. As noted above, however, the peak CH$_{3}$CN abundance, which occurs shortly after it thermally desorbs from grain surfaces, is not very sharp, and there is almost as much gaseous CH$_{3}$CN predicted at temperatures of $\sim$~200~K -- 300~K. For CH$_{3}$OCH$_{3}$, we observe \trot~$\approx$~100 -- 110~K, but the models predict \trot~$\geq$~200~K. This is a result of the fact that CH$_{3}$OCH$_{3}$ continues to be produced in the gas phase, after thermal desorption at approximately 70~K, via the reaction,
\begin{equation}
\label{e-de}
\rm CH_{3}OH_{2}^{+} + CH_{3}OH \to CH_{3}OCH_{4}^{+} + H_{2}O
\end{equation}
and subsequent dissociative recombination. The formation of CH$_{3}$OCH$_{3}$ is thus sensitive to the gas phase abundance of methanol which is over predicted in our models at its peak. The contribution of Reaction~\ref{e-de} to the CH$_{3}$OCH$_{3}$ abundance may, therefore, be over estimated in our models. We note that the predicted \tmax\ for C$_{2}$H$_{5}$CN agrees well with our measured \trot, but is much lower than 300~K, the temperature of the hot, compact component required to fit the HIFI emission (see Sec.~\ref{s-emis}), which is plotted as a green square in Figure~\ref{p-temp}. However, just as with CH$_{3}$CN, the peak abundance for C$_{2}$H$_{5}$CN is not sharp and similar abundances are predicted for temperatures $\gtrsim$~125~K.

The age of the hot core is estimated to be approximately 10$^{3}$--10$^{4}$~years based on measured D/H ratios of H$_{2}$O and NH$_{3}$ \citep{plambeck87, walmsley87, brown88}. The models presented in this study that have \twarm~=~2\sn$^{5}$~years, which produce better agreement overall with measured abundances of complex organics relative to models with \twarm~=~5\sn$^{4}$~years, require timescales $\gtrsim$~10$^{5}$~years to reproduce observed abundance levels (see Section~\ref{s-mod}). The time necessary to form complex organics in sufficient quantities according to our models is thus at least an order of magnitude larger than the age of the hot core as estimated by D/H ratios of lighter hydrides. Moreover, \citet{esplugues14}, who model the sulfur chemistry toward the Orion~KL hot core in detail, require an intermediate age of $\sim$5\sn$^{4}$~years to reproduce the observed abundances of S-bearing molecules in their best fitting models. We note that these timescales are not well constrained given the complex density/temperature structure and the uncertain dynamical history of Orion~KL. As a result, we do not consider our timescale of $\gtrsim$~10$^{5}$~years as a robust age estimate for the hot core or compact ridge. The fact that it is significantly longer than other hot core age estimates, however, suggests certain events, not modeled here, may have facilitated the production of complex molecules toward Orion KL early in its evolution. For example, if the gas in the hot core and compact ridge spent a significant amount of time at temperatures $\sim$~20 -- 40~K prior to warmup, the formation of complex organics may have been hastened due to the mobility of heavy radicals on grain surfaces at earlier times. Such a possibility is not unreasonable given that most molecules measured toward the extended ridge have measured rotation temperatures $\gtrsim$~20~K (Paper I). The clumpy structure of Orion~KL may also have accelerated the formation of complex organics because holes in dusty material could have provided pathways for UV photons to break up the molecular constituents of ice mantles, facilitating the formation of heavy radicals on dust grains. These caveats may also help to reconcile the under predictions of C$_{2}$H$_{3}$CN, C$_{2}$H$_{5}$CN, and CH$_{3}$OCHO by our model grid. We also note that if the hot core is indeed heated externally by an explosive event which occurred less than 720 years ago \citep{bally08, zapata09, zapata11, bally11, nissen12}, then the dynamical history of Orion~KL is far more complex than simulated here, which may also help to explain why some discrepancies exist between measured and predicted abundance levels.

\section{Conclusions}
\label{s-con}

Using spectral models fit to the molecular emission within the Orion~KL \herschel/HIFI spectral survey, we have identified the complex organics that are emitting in the hottest gas toward the different spatial/velocity components associated with this region. We find that the nitrogen bearing complex organics CH$_{3}$CN, C$_{2}$H$_{3}$CN, C$_{2}$H$_{5}$CN, and NH$_{2}$CHO systematically trace hotter gas than O-bearing complex species which contain no nitrogen: CH$_{3}$OH, C$_{2}$H$_{5}$OH, CH$_{3}$OCH$_{3}$, and CH$_{3}$OCHO. We emphasize that these results apply to molecules detected within a spatial/velocity component and are obtained using the emission predicted by our models, which reproduce the data well, as a direct diagnostic of the excitation. Hence, they do not depend on any model assumptions (e.g. LTE). If complex organics form on grain surfaces, these observations may indicate that complex N-bearing molecules are more difficult to remove from grain surfaces than organics which contain only H, C, and O. Another possibility is that the complex N-bearing species are forming via a high temperature gas phase chemistry and thus naturally trace hotter gas. 

We also compared our derived molecular abundances and rotation temperatures to chemical models derived using the OSU gas-grain chemical code, which includes both an isothermal collapse and warm-up phase. We find better agreement in the observed abundance levels of complex organics when \twarm\ is set to 2\sn$^{5}$~years as opposed to 5\sn$^{4}$~years. Abundances for C$_{2}$H$_{3}$CN, C$_{2}$H$_{5}$CN, and CH$_{3}$OCHO are under predicted for all models in our grid, though the model and observed abundances disagree by less than 3$\sigma$ for \twarm~=~2\sn$^{5}$~years. We also find that model \tmax\ values agree reasonably well (within 2$\sigma$) with observed rotation temperatures for all complex organics except CH$_{3}$CN and CH$_{3}$OCH$_{3}$. Finally, SO, SO$_{2}$, and OCS, i.e. S~$+$~O-bearing molecules, are among the molecules which trace the hottest gas toward Orion~KL. If these species are formed via gas phase reactions starting with the sublimation of H$_{2}$S from grains, our results may indicate that the formation routes of these species are most active at high temperatures. 

\acknowledgments
We gratefully acknowledge the anonymous referee for comments which improved this manuscript. HIFI has been designed and built by a consortium of institutes and university departments from across Europe, Canada and the United States under the leadership of SRON Netherlands Institute for Space Research, Groningen, The Netherlands and with major contributions from Germany, France and the US. Consortium members are: Canada: CSA, U.Waterloo; France: CESR, LAB, LERMA, IRAM; Germany: KOSMA, MPIfR, MPS; Ireland, NUI Maynooth; Italy: ASI, IFSI-INAF, Osservatorio Astrofisico di Arcetri-INAF; Netherlands: SRON, TUD; Poland: CAMK, CBK; Spain: Observatorio Astron\'{o}mico Nacional (IGN), Centro de Astrobiolog\'{i}a (CSIC-INTA). Sweden: Chalmers University of Technology - MC2, RSS \& GARD; Onsala Space Observatory; Swedish National Space Board, Stockholm University - Stockholm Observatory; Switzerland: ETH Zurich, FHNW; USA: Caltech, JPL, NHSC. HIPE is a joint development by the Herschel Science Ground Segment Consortium, consisting of ESA, the NASA Herschel Science Center, and the HIFI, PACS and SPIRE consortia. Support for this work was provided by NASA through an award issued by JPL/Caltech. E.~Herbst acknowledges funding from NASA/Jet Propulsion Laboratory to aid US participation in the \herschel/HIFI project.

\bibliographystyle{apj}
\bibliography{ms.bbl}

\clearpage

\begin{deluxetable}{lcccccccc}
\tablecaption{Model Predictions for Complex Molecules \label{t-obs}}
\tablecolumns{9}
\tablewidth{0pt}
\tabletypesize{\footnotesize}
\tablehead{\multicolumn{1}{c}{\tfinal}   &  \multicolumn{2}{c}{200~K}              &  \multicolumn{2}{c}{200~K}             &  \multicolumn{2}{c}{300~K}             &  \multicolumn{2}{c}{300~K}              \\
                  \multicolumn{1}{c}{\twarm} &  \multicolumn{2}{c}{5\sn$^{4}$~yr}  &  \multicolumn{2}{c}{2\sn$^{5}$~yr}  &  \multicolumn{2}{c}{5\sn$^{4}$~yr}  &  \multicolumn{2}{c}{2\sn$^{5}$~yr}   \\
\cline{1-9}
\colhead{Molecule} & \colhead{$n_{\rm x}$/$n_{\rm H_{2}}$}  & \colhead{\tmax\ (K)} & \colhead{$n_{\rm x}$/$n_{\rm. H_{2}}$} & \colhead{\tmax\ (K)} & \colhead{$n_{\rm x}$/$n_{\rm H_{2}}$} & \colhead{\tmax\ (K)} & \colhead{$n_{\rm x}$/$n_{\rm H_{2}}$} & \colhead{\tmax\ (K)}}
\startdata
CH$_{3}$CN              & 7.9\sn$^{-9}$  &  104    &  3.5\sn$^{-8}$   & 101    & 6.8\sn$^{-9}$  & 111     & 2.8\sn$^{-8}$ & 100   \\
C$_{2}$H$_{3}$CN    & 1.9\sn$^{-9}$  & 200    &  3.4\sn$^{-9}$   & 200    & 2.3\sn$^{-9}$   & 300    & 3.2\sn$^{-9}$ & 300   \\
C$_{2}$H$_{5}$CN    & 8.4\sn$^{-9}$  & 127    &  4.6\sn$^{-8}$   & 123    & 7.0\sn$^{-9}$   & 126    & 2.9\sn$^{-8}$ & 122   \\
NH$_{2}$CHO            & 2.3\sn$^{-6}$  & 200    &  4.1\sn$^{-6}$   & 200    & 3.1\sn$^{-6}$  & 300    & 5.0\sn$^{-6}$ & 295   \\

CH$_{3}$OH              &  4.8\sn$^{-5}$  & 119    &  4.0\sn$^{-5}$   & 115    & 4.9\sn$^{-5}$  & 118    & 4.1\sn$^{-5}$ & 122   \\
C$_{2}$H$_{5}$OH    &  5.0\sn$^{-8}$  & 127   &  1.8\sn$^{-7}$   & 123    & 3.8\sn$^{-8}$  & 126    & 1.5\sn$^{-7}$ & 122  \\
CH$_{3}$OCH$_{3}$ &  9.3\sn$^{-8}$  & 200   &  2.5\sn$^{-7}$   &  200   & 1.1\sn$^{-7}$  & 300    & 2.9\sn$^{-7}$ & 300  \\
CH$_{3}$OCHO         &  6.6\sn$^{-9}$  &  92    &  1.1\sn$^{-8}$   &  83     &  6.8\sn$^{-9}$  & 91    & 9.5\sn$^{-9}$ &  88  \\
\enddata
\tablecomments{Abundances represent peak values.}
\end{deluxetable}

\clearpage

\clearpage

\clearpage

\clearpage

\clearpage

\clearpage

\end{document}